# From Bench to Bedside: A DeepSeek-Powered AI System for Automated Chest Radiograph Interpretation in Clinical Practice

Yaowei Bai[1,2*], Ruiheng Zhang[3,4*], Yu Lei[2,5*], Jingfeng Yao[6], Shuguang Ju[7], Chaoyang Wang[8], Wei Yao[2,5], Yiwan Guo[2,5], Guilin Zhang[2,5], Chao Wan[9], Qian Yuan[10], Xuhua Duan[7], Xinggang Wang[6], Tao Sun[11], Yongchao Xu[3,4#], Chuansheng Zheng[2,5#], Huangxuan Zhao[1,4#], Bo Du[1,4#]

Affiliations:

[1] Department of Radiology, Union Hospital, Tongji Medical College, Huazhong University of Science and Technology, Wuhan, 430022, China; School of Computer Science, Wuhan University, Wuhan, 430072, China

[2] Hubei Provincial Clinical Research Center for Precision Radiology & Interventional Medicine, Wuhan 430022, China

[3] School of Computer Science, Wuhan University, Wuhan, 430072, China

[4] National Engineering Research Center for Multimedia Software and Hubei Key Laboratory of Multimedia and Network Communication Engineering, Wuhan University, Wuhan 430072, China

[5] Department of Radiology, Union Hospital, Tongji Medical College, Huazhong University of Science and Technology, Wuhan, 430022, China

[6] School of Electronic Information and Communications, Huazhong University of Science and Technology, Wuhan, 430074, China

[7] Department of Interventional Radiology, The First Affiliated Hospital of Zhengzhou University, Zhengzhou, 450000, China

[8] Department of Interventional Radiology, The First Affiliated Hospital of Henan University of Science and Technology, Luoyang, 471003, China

[9] Cancer Center, Union Hospital, Tongji Medical College, Huazhong University of Science and Technology, Wuhan, 430022, China.

[10] Department of Nephrology, Union Hospital, Tongji Medical College, Huazhong University of Science and Technology, Wuhan, 430022, China.

[11] Department of Interventional Radiology, The First Affiliated Hospital of University of Science and Technology of China, Hefei, 230001, China

*Contributed equally to the study;

#Corresponding authors. Email: yongchao.xu@whu.edu.cn; hqzcsxh@sina.com;

zhao_huangxuan@sina.com; dubo@whu.edu.cn

**Abstract**

A global shortage of radiologists has been exacerbated by the significant volume of chest X-ray workloads, particularly in primary care. Although multimodal large language models show promise, existing evaluations predominantly rely on automated metrics or retrospective analyses, lacking rigorous prospective clinical validation. Janus-Pro-CXR (1B), a chest X-ray interpretation system based on DeepSeek's Janus-Pro model, was developed and rigorously validated through a multicenter prospective trial (NCT06874647). Our system outperforms state-of-the-art X-ray report generation models in automated report generation, surpassing even larger-scale models including ChatGPT 4o (200B parameters), while demonstrating robust detection of eight clinically critical radiographic findings (area under the curve, AUC > 0.8). Retrospective evaluation confirms significantly higher report accuracy than Janus-Pro and ChatGPT 4o. In prospective clinical deployment, AI assistance significantly improved report quality scores (4.37 ± 0.50 vs. 4.11 ± 0.81, P < 0.001), reduced interpretation time by 18.5% (P < 0.001), and was preferred by a majority of experts (≥3 out of 5) in 52.7% of cases. Through lightweight architecture and domain-specific optimization, Janus-Pro-CXR improves diagnostic reliability and workflow efficiency, particularly in resource-constrained settings. The model architecture and implementation framework will be open-sourced to facilitate the clinical translation of AI-assisted radiology solutions.

**Introduction**

The global shortage of radiologists presents a critical challenge, with over 50% of the world's population lacking access to basic radiological diagnostic services[1,2]. This issue is particularly pronounced in low-income regions, where countries report just 1.9 radiologists per million residents, compared to 97.9 per million in high-income countries[3,4]. Chest X-ray (CXR), the most fundamental and widely utilized imaging modality, remains indispensable in clinical practices such as detecting pulmonary infections and screening for tumors. X-ray diagnostics contribute significantly to the workload of radiologists, especially in primary healthcare settings[5].

Recent advances in artificial intelligence (AI) demonstrate substantial potential to enhance diagnostic efficiency, optimize medical resource allocation, and maintain the quality of healthcare[6]. Currently, most AI applications focus on classifying and quantitatively analyzing imaging features for specific diseases, but clinical imaging diagnoses are far more complex than simple classification tasks[7]. Developing an intelligent imaging system with automated report generation at its core could be crucial in improving diagnostic efficiency and alleviating the strain on radiologists[8].

Despite the growing body of research on CXR report generation, most existing models are built from scratch[9-11], presenting inherent limitations such as low data efficiency, modal fragmentation, knowledge transfer challenges, and an exacerbation of the long-tail problem[12]. Transfer learning, leveraging pre-trained knowledge and cross-modal alignment, could significantly improve the accuracy and clinical relevance of report-generation systems[8,13]. However, many current models are either prohibitively large, underperforming, or non-open-source, limiting clinical applicability. Additionally, most prior studies have evaluated generated reports using natural language generation metrics, without assessing the actual clinical impact[14-17]. Though some studies have provided comprehensive evaluations[12,18,19], they have largely relied on retrospective data in collaboration scenarios between clinicians and AI, without prospective validation in real clinical settings. Consequently, the clinical value of multimodal large models in chest radiograph interpretation remains uncertain. The recently released open-source multimodal large model Janus-Pro by DeepSeek[20,21], with its combination of high performance and low cost, offers a new pathway for developing medical-specific report-generation systems. However, the application of this model in medical imaging has not been systematically tested, and existing general multimodal models lack task-specific optimization, necessitating further fine-tuning.

To address these gaps, this study introduces Janus-Pro-CXR, a lightweight CXR-specific model ([Figure 1](#)) developed from the unified Janus-Pro model and the public MIMIC-CXR medical imaging database[22] through supervised fine-tuning. With 1 billion parameters, the model achieves rapid imaging analysis with a latency of 1-2 seconds on a laptop equipped with a GeForce RTX 4060 (8GB). Its low fine-tuning cost further supports deployment in regions with limited medical resources. The model's performance in core tasks, including disease diagnosis and report generation, was rigorously evaluated using multicenter retrospective data from 27 hospitals. A multicenter prospective verification scheme was also implemented for clinical collaboration scenarios. Quantitative analysis of diagnostic efficacy and work efficiency in the collaborative mode between junior radiologists and AI revealed that AI assistance significantly improved report quality scores (4.37 ± 0.50 vs. 4.11 ± 0.81, $P < 0.001$), reduced interpretation time by 18.5%, and was preferred by a majority of experts (≥3 out of 5) in 52.7% of cases. The lightweight architecture and domain-specific optimization of Janus-Pro-CXR enhanced diagnostic accuracy and workflow efficiency, offering particular benefits for radiologists and settings with limited resources. The model architecture and implementation framework will be open-sourced to facilitate the clinical translation of AI-assisted radiology solutions.

**Results**

**Performance of Janus-Pro-CXR**

The general large model, Janus-Pro, underwent supervised fine-tuning using the MIMIC-CXR and CXR-27 datasets (multicenter retrospective data) ([Figure 2](#)). Baseline patient data are provided in [Supplementary Tables 1 and 2](#). In the retrospective study, comparative analyses between AI-generated reports and original reports were performed, evaluating automated report generation metrics, report quality scores, pairwise preference tests, agreement evaluations, and other relevant parameters as secondary outcomes.

In the assessment of automated report generation metrics, Janus-Pro-CXR demonstrated superior performance across multiple dimensions (Figure 3A and 3B, Supplementary Table 3). Evaluation on the MIMIC-CXR test set highlighted the model's exceptional ability to achieve a balance in disease recognition. Its macro-average evaluations (Macro-avg F1-14: 34.7/F1-5: 50.1) ranked highest among all models, confirming its capability for balanced recognition of both rare and common diseases. Notably, the model's Radiology Graph (RadGraph) score for the medical entity relationship extraction task reached 26.4, surpassing that of mainstream models. This advantage was further evident in the CXR-27 test set, where Janus-Pro-CXR excelled in natural language generation metrics. Its ROUGE-L (60.5) and BLEU-4 (44) scores were the highest among all models, validating the model's consistency with standard radiology reports. Additionally, the RadGraph score soared to 61.1, with both the micro-average (Micro-avg F1-14: 56.2) and macro-average (Macro-avg F1-14: 35.1) significantly outperforming similar lightweight models.

To assess consistency with the gold standard, Cohen's kappa coefficient was employed. The model performed exceptionally well in classifying diseases such as support devices and pleural effusion (Figure 3C). Diagnostic performance was also remarkable (Figure 3D, Supplementary Figure 1, and Supplementary Table 4), exhibiting excellent performance (AUC > 0.8) in detecting eight signs, including support devices (AUC = 0.967), pleural effusion (AUC = 0.927), and pneumothorax (AUC = 0.916). However, the model's ability to recognize complex or subtle signs, such as fractures and pleural thickening, requires further improvement. This may be due to relatively mild disease manifestations or significant variations in these signs. Additionally, the limited sample size of certain signs, such as edema and fracture, in the fine-tuning dataset may contribute to the model's suboptimal performance for these conditions.

**Clinical expert consensus on report quality**

Among the 1,317 cases in the CXR-27 test set, 300 images were randomly selected for subjective evaluation. First, a confusion matrix was used to assess the language professionalism and standardization of the reports generated by the large models, without considering content correctness. The confusion matrix results revealed that the reports generated by Janus-Pro-CXR closely resembled the style of published reports, making it difficult for evaluators to distinguish them from the original reports. In contrast, reports generated by Janus-Pro and ChatGPT 4o were easily identifiable (Figure 4A and Supplementary Figure 2). When comparing the reports generated by Janus-Pro-CXR, Janus-Pro, and ChatGPT 4o to the published reports, the evaluation team assigned a quality score of 3.22 ± 1.12 to the Janus-Pro-CXR-generated reports. This was significantly higher than the scores for reports generated by Janus-Pro (1.59 ± 0.62) and ChatGPT 4o (1.72 ± 0.75) (Figure 4B). The agreement scores between the Janus-Pro-CXR-generated reports and the published reports reached 3.10 ± 1.02, notably higher than the agreement scores for Janus-Pro (1.67 ± 0.60) and ChatGPT 4o (1.75 ± 0.75) (Figure 4C). In the pairwise preference test, 15.1% of the reports generated by Janus-Pro-CXR were favored by three or more evaluation experts, significantly surpassing the proportions for Janus-Pro (2.7%) and ChatGPT 4o (5.1%) (Figure 4D).

**Prospective validation of radiologist–AI collaboration**

The primary outcomes of this study included report quality scores, pairwise preference tests, agreement evaluations, and reading time for the radiology reports produced by the two groups of junior radiologists in the prospective study. In the prospective study, junior radiologists in the AI-assisted group used reports generated by Janus-Pro-CXR as references and modified the content as needed, while radiologists in the SCP group independently drafted their reports. For each patient's imaging data, two junior radiologists from different groups independently reviewed the scans and generated separate reports. These reports were then reviewed, revised, and finalized by a senior radiologist before being issued for clinical use.

This prospective study demonstrated that AI significantly enhanced the quality of radiology reports created by junior radiologists (Table 1). In terms of report quality, the AI-assisted group achieved a mean score of 4.37 ± 0.50, significantly higher than the standard clinical practice (SCP) group's score of 4.11 ± 0.81 ($\Delta$ = 0.24, P < 0.001, 95% CI = 0.160-0.328) (Figure 4E). The agreement score of the AI-assisted group's reports, assessed using the RADPEER scoring system, was 4.27 ± 0.60, notably higher than the SCP group's score of 4.09 ± 0.86 ($\Delta$ = 0.18, P < 0.001, 95% CI = 0.089-0.279) (Figure 4F). In the pairwise preference test between radiologists and AI, 30% of the reports generated by AI were preferred by three or more evaluation experts. When comparing radiologists with and without AI assistance, 52.7% of AI-assisted reports were favored by three or more experts (Figure 4G, 4H, and 4I). Examples illustrating the improvement in report quality through AI assistance are presented in Supplementary Table 5. The analysis of work efficiency showed that the AI-assisted group wrote reports in an average of 119.4 ± 45.3 seconds, significantly faster than the SCP group, which took 146.6 ± 49.8 seconds (Δ = 27.2s, 18.5% reduction, $P < 0.001$, 95% CI = 19.47-34.83) (Figure 4J). Subgroup analysis revealed that for complex cases (those with ≥ 3 imaging manifestations), the time advantage for the AI-assisted group remained substantial (142.1 ± 18.1 vs. 181.1 ± 22.4 seconds, Δ = 39.0s, 21.5% reduction, $P < 0.001$, 95% CI = 38.20-39.79).

**Discussion**

In this study, Janus-Pro-CXR, an intelligent report generation system tailored for radiology clinical practice, was developed based on the DeepSeek multimodal model (Janus-Pro). The domain-specific optimization of Janus-Pro-CXR has yielded substantial advantages in natural language generation metrics[12,23-26] and demonstrated outstanding performance in diagnostic accuracy and report generation quality. Prospective clinical validation confirmed that Janus-Pro-CXR not only significantly enhances the diagnostic accuracy of radiologists but also markedly improves their work efficiency. Furthermore, with just 1 billion parameters, the model achieves state-of-the-art performance in automated report generation metrics, outperforming both specialized large models in its class and higher-parameter models including ChatGPT 4o (200B)[27]. This parameter efficiency gives it a distinct advantage for deployment in resource-constrained settings.

The lightweight architecture of the model plays a critical role in facilitating the clinical application of AI-assisted diagnostic technologies[28]. In the present study, Janus-Pro-CXR, a lightweight CXR-specific model derived from the unified Janus-Pro, was introduced through supervised fine-tuning. Despite having 1 billion parameters, the model achieves rapid imaging analysis with a latency of 1-2 seconds on a laptop equipped with a GeForce RTX 4060 (8GB). This characteristic is crucial for practical deployment in primary healthcare settings. Unlike traditional AI models that require high-performance computing hardware, Janus-Pro-CXR maintains excellent diagnostic performance while significantly lowering the hardware requirements, making AI-assisted diagnostic services accessible in resource-constrained settings[29]. Notably, models often require further fine-tuning when applied to different populations. The model requires only 6,000 images for domain adaptation fine-tuning to reach professional-level diagnostic accuracy, and its low fine-tuning cost further enhances its applicability. The lightweight design does not compromise the model's performance; rather, through domain-specific optimization, it surpasses both general and specialized large models in CXR diagnosis tasks, achieving exceptional results in key metrics, such as RadGraph score and F1 values for disease recognition.

Despite the increasing number of studies on AI-based report generation, which holds great potential for optimizing radiology workflows, there remains a lack of prospective validation of automatic report generation systems in real-world clinical scenarios. This gap may be attributed to the limited accuracy of reports generated by current systems and the absence of a widely accepted framework for evaluating clinical utility. Building upon the retrospective study that assessed the accuracy of Janus-Pro-CXR report generation, this study uniquely conducted a multicenter prospective clinical validation. The results revealed that Janus-Pro-CXR significantly enhanced the diagnostic quality of junior radiologists, improving report quality scores by 0.24 points, while also reducing the average report time by 27.2 seconds (18.5% reduction). Furthermore, the positive rate of pneumonia diagnosis in the AI-assisted group was notably higher than in the SCP group (55.2% vs. 36.4%, $P <$ 0.001). The AI system likely enhanced the decision-making confidence of junior radiologists by offering structured descriptions of abnormal signs and diagnostic prompts. In cases of uncertainty, radiologists in the AI-assisted group were able to convert ambiguous observations (e.g., "possible bilateral bronchitis") into definitive diagnoses (e.g., "bilateral bronchitis"). This confidence boost is closely linked to the diagnostic capabilities and clinical experience of junior radiologists[30].

AI assistance may also enhance the work efficiency of senior physicians. The generation and release of radiology diagnostic reports adhere to a stringent standardized workflow. Initially, junior radiologists perform comprehensive imaging analysis to generate preliminary reports, which subsequently undergo dual validation by senior radiologists to ensure diagnostic accuracy and protocol compliance. Upon verification, the reports are electronically signed by senior radiologists for official release, with real-time synchronization to the hospital information system for clinical reference. Notably, the AI-assisted workflow not only enhances the efficiency of junior radiologists but also potentially supports senior radiologists' review process by providing clearer diagnostic suggestions. This may reduce the need for report modifications by senior physicians and accelerate the review speed, thereby improving overall workflow efficiency.

This study has several limitations. Firstly, the model's recognition performance for complex or subtle signs, such as edema and pleural thickening, is relatively constrained, a common challenge with current large models[31]. This limitation may be due to the mild manifestations and variable morphologies of these signs, as well as insufficient sample representation of some conditions in the training and fine-tuning datasets. However, AI collaboration has still contributed to improvements in report quality and work efficiency. Future research should focus on optimizing the model's ability to recognize subtle signs and ensuring that training and fine-tuning datasets are more balanced and clinically representative. Furthermore, in actual clinical practice, radiologists have access to a comprehensive review of patients' medical histories and prior imaging results, which plays a critical role in improving the accuracy of imaging diagnoses[32]. In the present study, however, the model's fine-tuning phase did not incorporate such clinical background information. Moving forward, future studies should integrate comprehensive clinical data into the system to enable more accurate and thorough radiological disease diagnosis. Additionally, while this study provides preliminary clinical validation, the prospective validation was conducted in a limited number of hospitals. However, the model's technical advantages—such as its low parameter count and fast convergence—facilitate efficient fine-tuning and rapid deployment across new institutional data. A large-scale, multi-center clinical study will be conducted to further evaluate the model's generalizability. Finally, the collaboration mode between clinicians and AI can be more intricate than what was explored in this study. The ideal form of interaction should involve two-way real-time communication, similar to a senior consultation expert who not only answers radiologists' questions promptly but also offers proactive suggestions for diagnostic reports, such as identifying potential misdiagnoses or missed lesions. Although recent research on multimodal interactive medical AI has laid the foundation for such collaborative diagnostic systems[33,34], much work remains to develop an intelligent report assistance system that fully meets clinical needs.

This study represents the first prospective validation of Janus-Pro-CXR, highlighting the clinical value of AI-radiologist collaboration in real-world diagnostic practice. Through systematic supervised fine-tuning and comprehensive clinical validation, Janus-Pro-CXR not only improved diagnostic accuracy and workflow efficiency but also successfully transitioned from an auxiliary tool to a reliable "digital colleague." This advancement has proven particularly beneficial for radiologists and healthcare providers, especially in resource-limited settings. Although the current validation focused on CXR interpretation, the Janus-Pro-CXR framework offers a scalable paradigm with potential applications across various imaging modalities, including CT, MRI, and ultrasound. The core technological advantages of this system—its lightweight architecture, domain-specific optimization, and seamless human-AI collaborative workflow—demonstrate broad applicability across medical imaging domains.

**Methods**

**Ethics Approval and Consent**

Approval for this study was granted by the Institutional Review Boards of all participating centers, and it was duly registered with ClinicalTrials.gov (Identifier: NCT06874647). Informed consent was waived for the retrospective studies, while it was properly obtained for the prospective trials. All trials adhered to the CONSORT-AI and SPIRIT-AI reporting guidelines[35-37].

**Model Establishment**

The study utilized Janus-Pro (1B) as the base model, which was fine-tuned using Low-Rank Adaptation (LORA)[38] with a rank of 64. Fine-tuning was performed in two phases: first with the MIMIC-CXR dataset to acquire proficiency in basic diagnosis and report generation, and subsequently with the CXR-27 dataset to address data variability and adapt the writing style. The detailed process for model development is outlined in Supplementary Appendix 1. Regarding the interface of the model, relevant information can be obtained from Supplementary Figure 3.

**Trial design**

This research encompassed both retrospective and prospective components. The retrospective study focused on fine-tuning and evaluating the large model, while the prospective study aimed to explore AI collaboration with radiologists. In the retrospective study, 166,025 data points from the MIMIC-CXR dataset (Supplementary Appendix 2) and 5,267 records from 27 hospitals in China (CXR-27 dataset) were used for supervised fine-tuning. The remaining data from the CXR-27 dataset (n = 1,317) and a portion of the MIMIC-CXR dataset (n = 1,471) were used to assess model performance. The prospective study was conducted across three medical institutions in China (Supplementary Appendix 3).

**Participants**

Following data cleaning and quality screening of the MIMIC-CXR dataset (n=222,758), 167,496 radiographic images met inclusion criteria. Of these, 166,025 images were allocated for first-stage model fine-tuning, while 1,471 images were designated as the test set. For the retrospective study component, we obtained posteroanterior chest radiographs from 6,747 patients. Subsequent screening yielded 6,584 qualifying images, with 5,267 utilized for second-stage model fine-tuning and 1,317 reserved for final testing. The inclusion criteria for retrospective data were: 1) clear and complete chest X-ray images, and 2) definitive imaging diagnosis conclusions; the exclusion criteria were: 1) CXR images with severe quality issues, and 2) incomplete clinical medical records. The baseline characteristics of patients in the retrospective study are presented in Supplementary Table 1. For the prospective study, data were sourced from three hospitals in China, with a total of 297 patients enrolled in the AI-radiologist collaboration. The baseline characteristics for the prospective study are provided in Supplementary Table 2. For prospective clinical validation, the inclusion criteria included: 1) clinically suspected thoracic diseases (such as pneumonia, tuberculosis, or lung cancer) requiring CXR-assisted diagnosis, 2) patients providing written informed consent for research data use, and 3) complete clinical records (including chief complaints, medical history, and laboratory test results); the exclusion criteria were: 1) substandard CXR image quality (including severe motion artifacts, over-/underexposure, or missing anatomical structures), and 2) pregnant or lactating women.

**Automated report generation metrics evaluation of report quality**

The fine-tuned model, Janus-Pro-CXR, was evaluated using both automated report generation metrics and subjective indicators. For automated report generation, the evaluation included the RadGraph F1 score, as well as BLEU (1-gram/4-gram) and ROUGE-L scores. Additionally, the CheXbert[39] and DeepSeek annotation tools (Supplementary Appendix 4) were employed to calculate the micro-average and macro-average F1 scores for 14 types of diseases (F1-14) and the top 5 most common diseases (F1-5), assessing the model's ability to identify key imaging manifestations. Detailed methods for these evaluations are outlined in Supplementary Appendix 5.

**Subjective Evaluation of Report Quality**

Subjective evaluation was performed by a team of five radiologists, each with 8 to 15 years of experience. All evaluators received the necessary training before participating in the evaluation. For each case, the evaluators were blinded to which report was generated by the large model. A total of 300 cases were randomly selected from the retrospective CXR-27 test set for subjective assessment. The following evaluation methods were used: (1) Confusion Matrix: Without considering content accuracy, the order of reports generated by the large model and the published reports was randomized. The confusion matrix was then used to assess the professionalism of medical terms, logical coherence, diagnostic suggestions, and other report aspects. (2) Report Quality Score: Evaluators rated the large model's report generation capability against the published imaging reports, using a five-point Likert scale (5 representing the highest quality, 1 the lowest). The specific criteria for scoring are provided in Supplementary Appendix 6. (3) Pairwise Preference Test: Each evaluator selected the superior report between those generated by the large model and the published reports. (4) Agreement Evaluation: The RADPEER scoring system[40] was employed to assess the agreement between the original and generated reports, including the clinical significance of any differences. Detailed RADPEER scoring criteria are found in Supplementary Appendix 7.

**Evaluation of the collaboration between AI and radiologists**

The generation and release of radiology diagnostic reports followed a stringent standardized workflow. Junior physicians conducted a thorough review and analysis of the imaging data to draft initial diagnostic reports. These reports were then reviewed by senior physicians for accuracy and compliance with diagnostic standards. Upon verification, the reports were electronically signed by senior physicians and officially released, with real-time synchronization to the hospital information system for clinical use. Patients could access their diagnostic results through various channels, including hospital apps, self-service terminals, or printed reports. In the prospective study, six radiologists were randomly assigned to either the AI-assisted group or the SCP group in a 1:1 ratio. Randomization was conducted using a computer-generated sequence with sealed envelopes. In the AI-assisted group, junior radiologists used the reports generated by the large model as a reference and had the option to modify them as needed. In contrast, radiologists in the SCP group completed the reports independently. For each patient, two junior radiologists from different groups independently reviewed the images and produced two separate reports. These reports were then reviewed and revised by senior radiologists before being finalized for patient delivery. The quality of reports generated by the two groups of junior radiologists was assessed using report quality scores, pairwise preference tests, and agreement evaluations. Additionally, the time taken to write reports (reading time) was compared to evaluate the impact of the large model on work efficiency. Reading time was defined as the period from when the radiologists began examining the CXR until the final report was completed. This time was automatically recorded by the system. During the image interpretation process, junior radiologists had access to the full clinical medical history and prior imaging results of the patients in real time. To simulate real clinical conditions, key patient medical history information was incorporated into the Prompts for the large language model.

**Outcomes**

The primary outcomes of this study included report quality scores, pairwise preference tests, agreement evaluations, and reading time for the radiology reports produced by the two groups of junior radiologists in the prospective study. Secondary outcomes comprised the automated report generation metrics, report quality scores, pairwise preference tests, agreement evaluations, and other relevant metrics comparing the original reports with those generated by the model in the retrospective study.

**Statistical analysis**

Sample size determination for the prospective study was based on the pre-experiment results regarding report quality scores (power=0.90, Supplementary Appendix 8). Independent sample t-tests were conducted to compare differences between the two groups, while one-way analysis of variance (ANOVA) was used to assess differences across multiple groups. The model's classification performance for various chest diseases was evaluated using the ROC curve and the AUC, with probability thresholds ranging from 0 to 1. Cohen's kappa coefficient was employed to assess the agreement between the model's results and the gold standard. Kappa values range from -1 to 1, with a value of 1 indicating perfect agreement, 0 representing agreement due to chance, and -1 signifying perfect disagreement. The Kendall's W concordance coefficient was calculated to evaluate the consistency of results across multiple raters. All statistical analyses were performed using two-sided tests, with a significance level set at $P < 0.05$. Statistical analysis was performed using R software (version 4.2.1).

**Acknowledgements:** We extend our sincere gratitude to Dr. David Ouyang, MD (Assistant Professor, Department of Cardiology, Division of AI in Medicine, Cedars-Sinai Medical Center, Los Angeles, California, USA) for his invaluable feedback and insightful suggestions during the revision of this manuscript.

**Data availability statement:** Part of retrospective data are publicly available (https://github.com/ZrH42/Janus-Pro-CXR), the remaining data is available for research purposes upon reasonable request from the corresponding author, Huangxuan Zhao.

**Code availability statement:** The codes used can be available from the URL (https://github.com/ZrH42/Janus-Pro-CXR).

**Contributions**

B.Y., Z.R., L.Y., X.Y. Z.C., Z.H and D.B. had major involvement in study conception or design. Y.J., J.S., W.C., Y.W., G.Y., Z.G., W.C. Y.Q., D.X., W.X. and S.T. had substantial involvement in data acquisition.

B.Y. and L.Y. had major involvement in data analysis. B.Y., Y.W., G.Y. and Z.G. had major involvement in data interpretation. All authors had access to the data and participated in writing, reviewing, and revising the manuscript. All authors approved the final manuscript for publication. All authors accept responsibility for the accuracy and integrity of all aspects of the research.

**Conflict of Interest Disclosures:** The authors declared no conflicts of interest.

**Funding/Support:** This work was funded by the National Natural Science Foundation of China (Grant No. 82472070) and the National Key Research and Development Program of China (Grant No. 2023YFC2413500).

**Table 1.** **Subjective scores of the retrospective study and the prospective study.**

| | | Retrospective | | | Prospective | |
|---|---|---|---|---|---|---|
| | | Janus-Pro | ChatGPT 4o | Janus-Pro CXR | SCP | AI-assisted |
| Report Quality Score | Score | 1.59±0.62 | 1.72±0.75 | **3.22±1.12** | 4.11±0.81 | **4.37±0.50** |
| | Kendall's W | 0.626 | 0.720 | 0.719 | 0.736 | 0.774 |
| | P value* | <0.001 | <0.001 | <0.001 | <0.001 | <0.001 |
| Agreement Score | Score | 1.67±0.60 | 1.75±0.75 | **3.10±1.02** | 4.09±0.86 | **4.27±0.61** |
| | Kendall's W | 0.583 | 0.677 | 0.598 | 0.631 | 0.574 |
| | P value* | <0.001 | <0.001 | <0.001 | <0.001 | <0.001 |

*The marked P value is the result of Kendall's W concordance coefficient test (to evaluate the inter-rater consistency). The strength of agreement between the observers was classified as poor (W＜0.2), fair (0.2 ≤W＜0.4), moderate (0.4 ≤W＜0.6), strong (0.6 ≤W＜0.8), or super (0.8 ≤W＜1.0). Bold indicates the highest score result.

**Figures and figure legends**

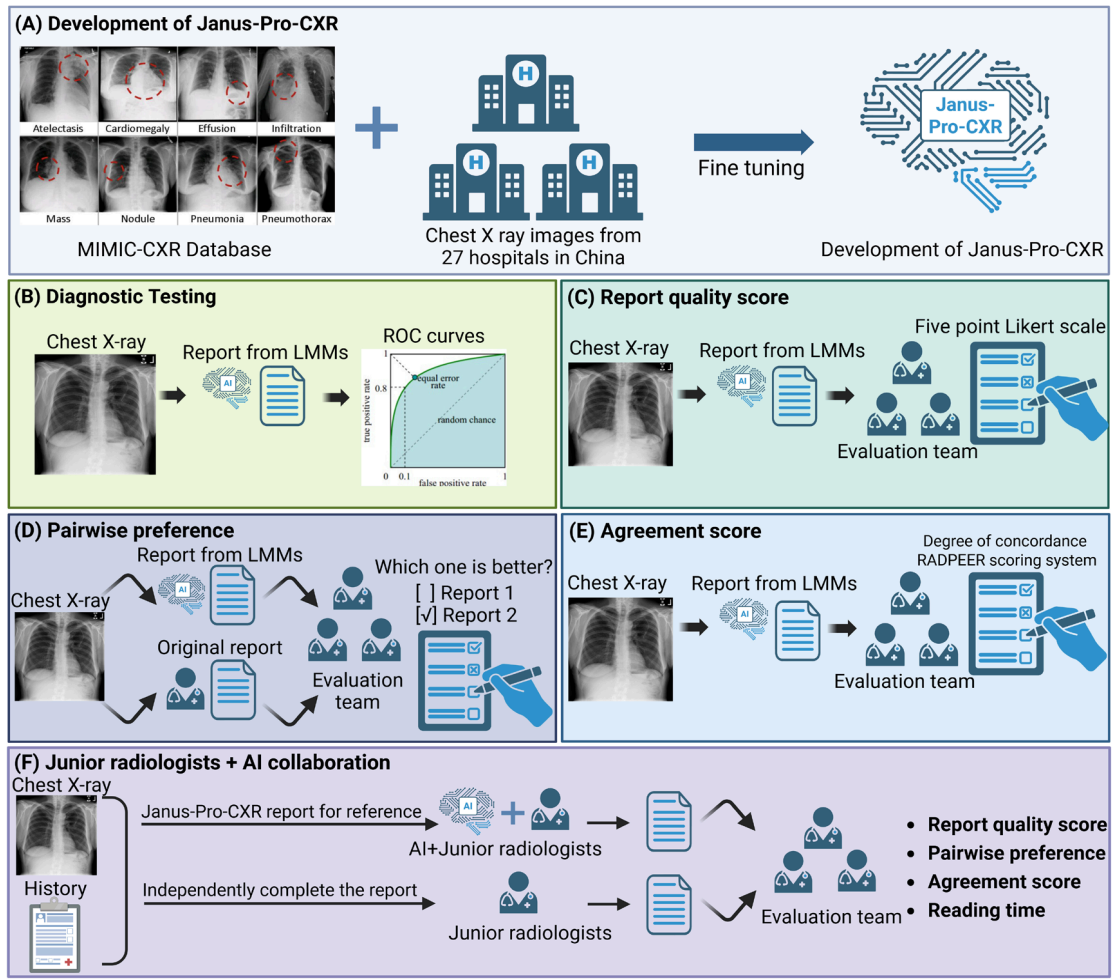

**Figure 1.** Flow chart of the systematic evaluation in this study. (A) Janus-Pro-CXR was constructed through supervised fine-tuning using the MIMIC-CXR dataset and the multicenter retrospective dataset (CXR-27). The retrospective data were sourced from 27 medical centers in China, collectively referred to as CXR-27. (B) The model's performance was assessed using automated report generation metrics. The quality of the generated reports was evaluated through report quality scores (C), preference tests (D), and agreement evaluations (E). (F) The clinical utility of the model was evaluated through clinical collaboration with radiologists. The figure was created using BioRender (https://biorender.com/).

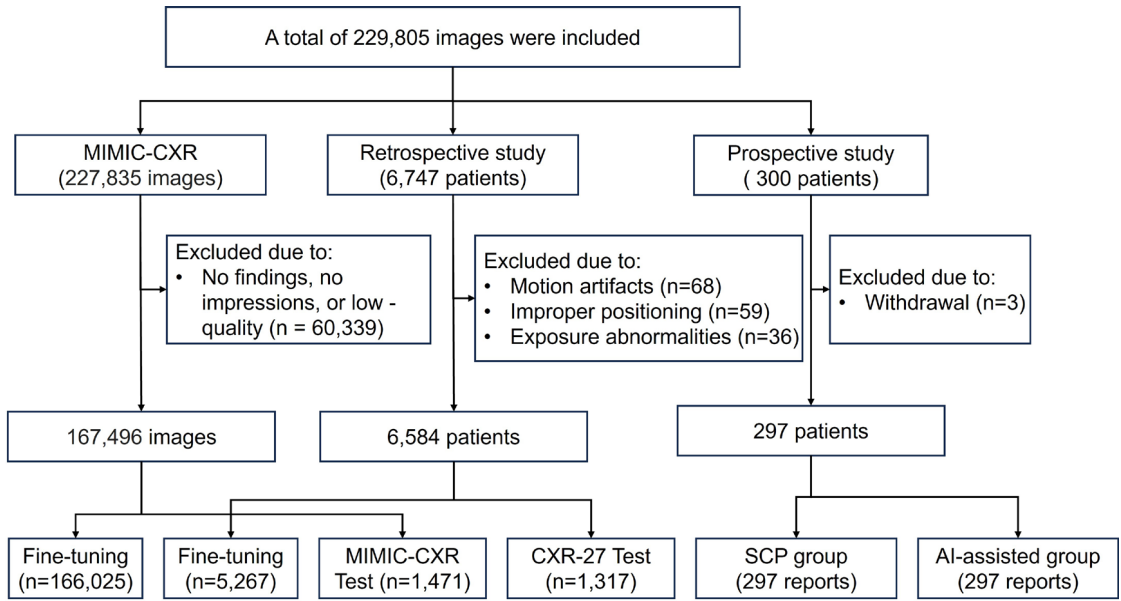

**Figure 2.** Flow chart of patient inclusion and exclusion. Janus-Pro-CXR was constructed through supervised fine-tuning using the MIMIC-CXR dataset and the multicenter retrospective dataset (CXR-27). The retrospective data were sourced from 27 medical centers in China, collectively referred to as CXR-27.

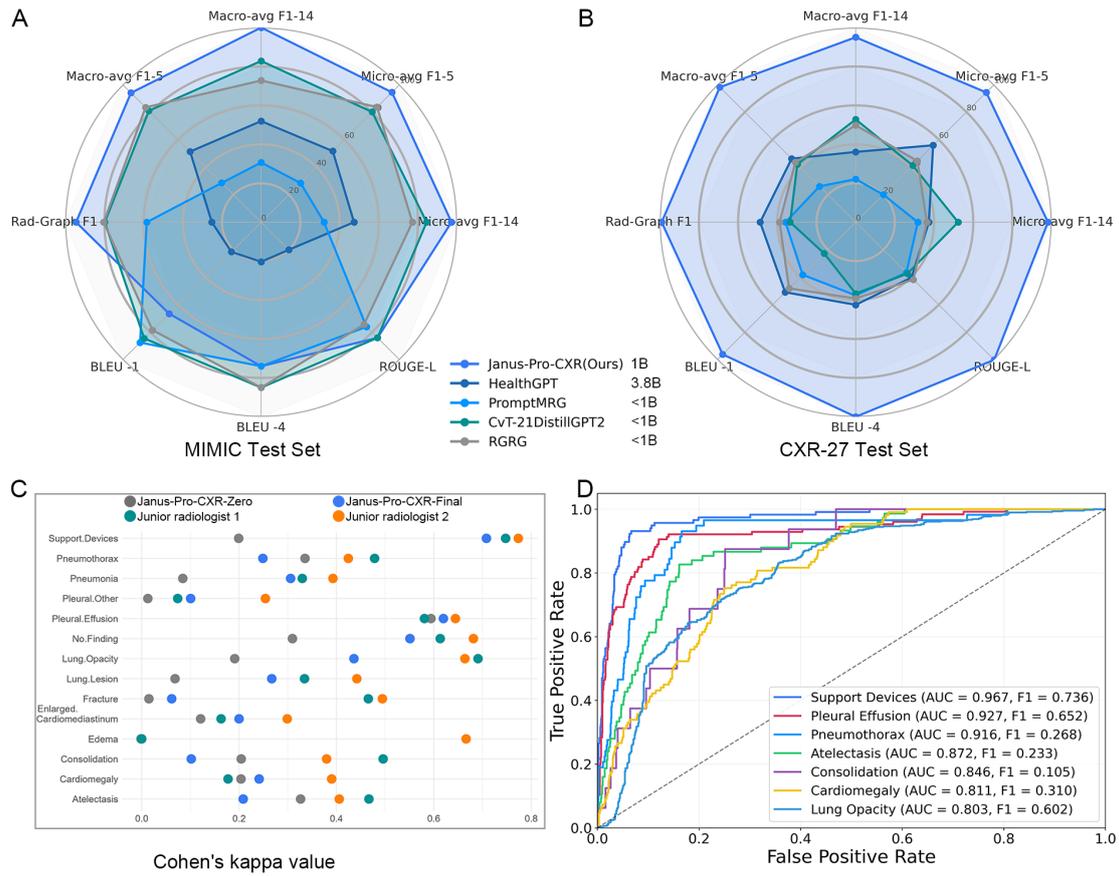

**Figure 3.** Performance of Janus-Pro-CXR on automated report generation metrics for the MIMIC-CXR test set (A) and the CXR-27 test set (B). (A) Automated report generation metrics for the MIMIC-CXR test set. Annotation was performed using the CheXbert labelling tool (uncertain labels treated as positive). The same open-source large model was tested with parameter configurations consistent with the published research. The top 5 diseases are: Atelectasis, Cardiomegaly, Edema, Consolidation, and Pleural Effusion. (B) Automated report generation metrics for the CXR-27 test set. Annotation was performed using the DeepSeek labelling tool, and the open-source large model was tested with consistent parameter configuration. The top 5 diseases include: Support Devices, Pleural Effusion, Lung Opacity, Pneumonia, and Lung Lesion. (C) Cohen's kappa coefficient for evaluating consistency between the model and the gold standard. Janus-Pro-CXR-Zero refers to the model fine-tuned with the MIMIC-CXR dataset, while Janus-Pro-CXR-Final indicates the model fine-tuned with the CXR-27 dataset, building on Janus-Pro-CXR-Zero. (D) Classification performance of the model for various chest diseases. Evaluated through the receiver operating characteristic (ROC) curve and the area under the curve (AUC).

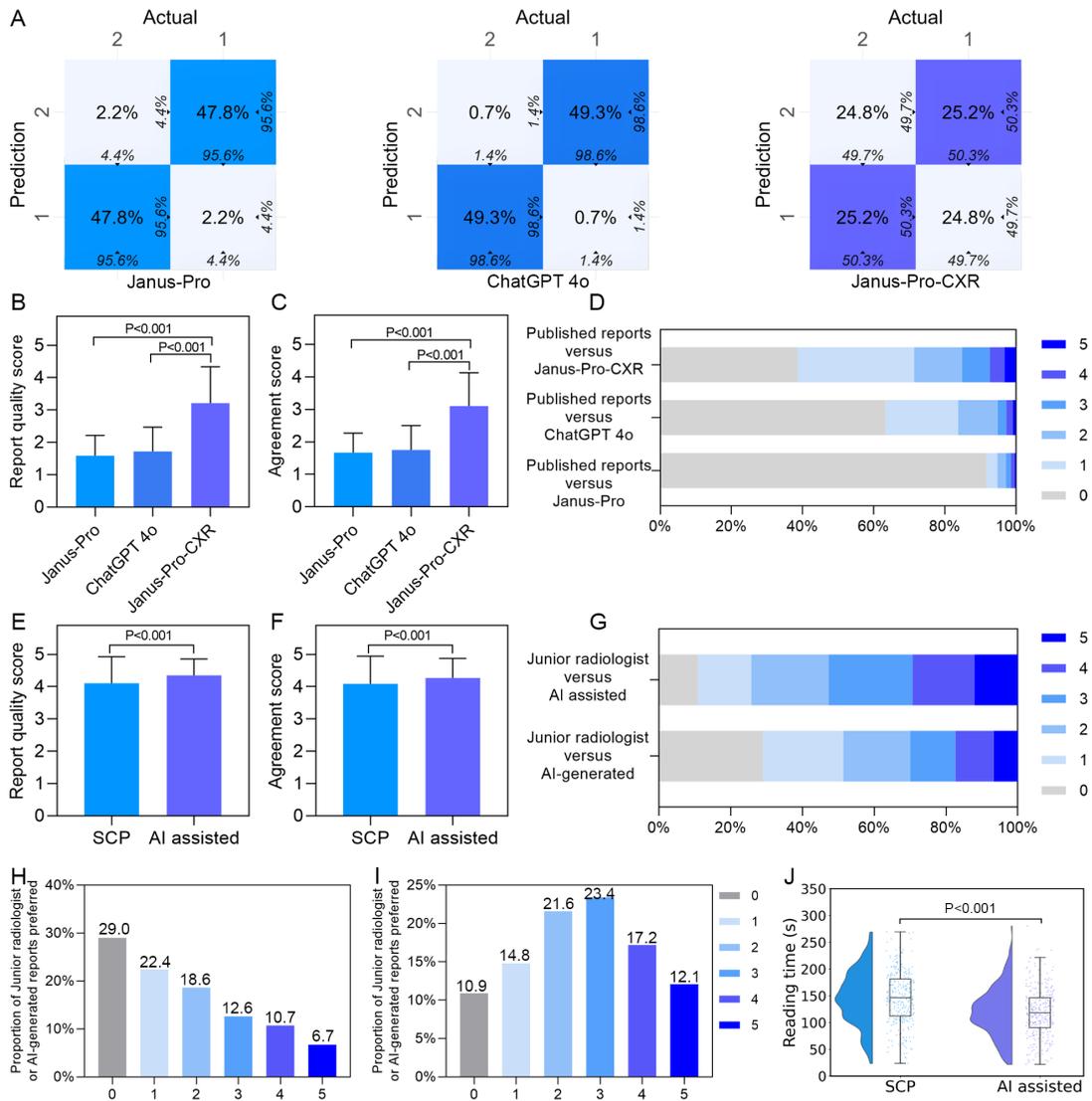

**Figure 4.** Subjective evaluations of the retrospective and prospective studies. (A) Confusion matrix for evaluators to identify reports generated by large models in the retrospective study. (B) Report quality scores of the three large models in the retrospective study. (C) Report agreement scores of the three large models in the retrospective study. (D) Preference tests comparing published reports with those generated by Janus-Pro-CXR, ChatGPT 4o, and Janus-Pro in the retrospective study. (E) Report quality scores for the SCP group and the AI collaboration group in the prospective study. (F) Report agreement scores for the SCP group and the AI collaboration group in the prospective study. (G) Preference tests comparing reports generated by junior radiologists and those generated by AI, as well as those generated by junior radiologists with AI collaboration, in the prospective study. (H) Preference test comparing reports generated by junior radiologists and AI. (I) Preference test comparing reports generated by junior radiologists with AI collaboration. (J) Reading time for the SCP group and the AI collaboration group in the prospective study.

**From Bench to Bedside: A DeepSeek-Powered AI System for Automated Chest Radiograph Interpretation in Clinical Practice**

# Contents

**Supplementary Appendix 1** Model establishment

This study used Janus-Pro (1B) as the base model. We employed LORA (Low Rank Adaptation) to fine-tune the large model and set the rank of LORA to 64.

We adjusted the long side of all chest X-ray images and filled the short side with a background color (RGB: 127, 127, 127), and finally adjusted them to a size of 384×384 pixels. In addition, we processed all reports into the format of "FINDINGS" + "IMPRESSION".

Our model was fine-tuned twice, first with the MIMIC-CXR dataset and then with the CXR-27 dataset. Both fine-tuning processes were completed on 4 Nvidia RTX 3090 (24GB) graphics cards. The first-stage fine-tuning aimed to enable the model to master basic visual diagnosis and report generation capabilities. It was iterated for 35K steps (step) on the cleaned MIMIC-CXR dataset with a batch size of 128, taking approximately 72 hours. The second-stage fine-tuning aimed to make the model adapt to data differences and learn the report writing style. It was iterated for 0.4K steps (step) on the CXR-27 dataset with a batch size of 128, taking approximately 1 hour.

In our experiment, all tests were completed on 1 Nvidia RTX 3090 (24GB) graphics card, and the Natural Language Generation (NLG) metrics and clinical metrics of the model were comprehensively evaluated on the test dataset.

**Supplementary Appendix 2** MIMIC-CXR dataset

MIMIC-CXR (Medical Information Mart for Intensive Care - Chest X-Ray) is a large-scale publicly available medical imaging dataset containing over 370,000 chest X-ray images in DICOM format and 227,835 corresponding radiology reports in text format, covering imaging data from intensive care unit (ICU) patients[1]. The radiology reports, which are divided into "Findings" and "Impression" sections, have undergone de-identification to protect patient privacy. The dataset includes the following diagnostic labels: No Finding (healthy), Enlarged Cardiomediastinum, Cardiomegaly, Lung Opacity, Lung Lesion, Edema, Consolidation, Pneumonia, Atelectasis, Pneumothorax, Pleural Effusion, Pleural Other, Fracture, and Support Devices.

**Supplementary Appendix 3** Prospective medical centers

Union Hospital, Tongji Medical College, Huazhong University of Science and Technology; The First Affiliated Hospital of Zhengzhou University; The First Affiliated Hospital of Henan University of Science and Technology

**Supplementary Appendix 4** Automated labeling tool for chest x-ray reports

To ensure a fair comparison with previously published research results on the MIMIC-CXR dataset, we used the CheXpert labeling tool[2] to extract binary classification labels (indicating radiographic findings) from radiology reports to evaluate our model's performance.

To more accurately extract binary classification labels for radiographic findings, we developed an automated annotation tool based on DeepSeek for extracting structured labels of 14 key clinical findings (such as pneumonia, pleural effusion, etc.) from chest X-ray reports. The tool analyzes free-text reports by calling the deepseek-chat model and outputs JSON-formatted results indicating the presence or absence (0/1) of each finding, with support for batch processing of CSV files. The tool incorporates error-handling mechanisms and employs incremental saving to ensure data processing reliability. This tool significantly improves dataset labeling efficiency and provides standardized labels for model training and validation. A comparison of the F1 scores between the two labeling tools is as follows:

| Label | Chexpert_F1 | DS_F1 |
|---|---|---|
| Enlarged Cardiomediastinum | 0.1197 | 0.9524 |
| Cardiomegaly | 0.5434 | 1.0000 |
| Lung Opacity | 0.6414 | 0.9959 |
| Lung Lesion | 0.6361 | 0.9837 |
| Consolidation | 0.3333 | 0.9677 |
| Pneumonia | 0.1255 | 0.9920 |
| Atelectasis | 0.4330 | 1.0000 |
| Pneumothorax | 0.8627 | 1.0000 |
| Pleural Effusion | 0.8108 | 1.0000 |
| Pleural Other | 0.5468 | 0.9948 |
| Fracture | 0.5909 | 0.9811 |
| Support Devices | 0.9511 | 0.9957 |
| No Finding | 0.1391 | 0.9977 |

**Supplementary Appendix 5** Multi-dimensional automated report generation metrics to evaluate report generation quality

We employed a multi-dimensional automated metric system to evaluate report generation quality, encompassing both clinical finding recognition accuracy and text generation quality. Using CheXbert and DeepSeek annotation tools, we calculated micro-averaged/macro-averaged F1 scores for 14 disease categories (F1-14) and the top 5 most common diseases (F1-5) to assess the model's capability in identifying key radiographic findings. RadGraph F1 scores were adopted to quantify the accuracy of anatomical structure descriptions and their relationships. BLEU (1-gram/4-gram) and ROUGE-L metrics were used to measure lexical and semantic similarity between generated reports and reference reports. All comparative experiments followed the original paper's parameter configurations, with open-source models tested directly and non-open-source models citing literature data.

**Supplementary Appendix 6** Radiology report quality scoring criteria (5-point likert scale)

| Score | Criteria |
|---|---|
| 5 | 1. Flawless structure, terminology, and diagnostic reasoning<br>2. All findings precisely described with clinical context<br>3. Diagnosis matches gold standard perfectly |
| 4 | 1. Minor formatting issues or slight terminology deviations<br>2. 1-2 non-critical findings described less precisely<br>3. Diagnosis correct but may lack nuance |
| 3 | 1. Basic sections present (findings/impression)<br>2. Some findings unclear/partially wrong, but main abnormalities identified<br>3. Diagnosis directionally correct |
| 2 | 1. Missing major sections or confusing flow<br>2. Multiple errors affecting interpretation<br>3. Diagnosis questionable but not dangerous |
| 1 | 1. Unreadable or medically dangerous errors<br>2. Critical findings missed/misrepresented<br>3. Diagnosis contradicts imaging |

**Supplementary Appendix 7** RADPEER scoring criteria

The RADPEER system, established by the American College of Radiology (ACR), is a peer review program designed to evaluate the interpretation accuracy of radiologists. Using the RADPEER scoring system, the degree of concordance is assessed with discrepancies and agreements graded according to the following criteria:

- **Category 1**: Clinically significant discrepancy

- **Category 2**: Clinically insignificant discrepancy

- **Category 3**: Understandable clinically significant miss (excusable given case complexity)

- **Category 4**: Understandable clinically insignificant miss (excusable due to subtle findings, case complexity, or diagnostic difficulty)

- **Category 5**: Complete agreement

**Supplementary Appendix 8** Sample size calculation

Based on preliminary trial data (n=30), the mean difference in imaging report scores between the AI-assisted group and the SCP group was 0.207 points (standard deviation σ_diff=0.872). To detect this difference (two-tailed α=0.05, power=0.90) while accounting for a 30% dropout rate, the calculated required sample size was 270 cases. To ensure robust results capable of detecting statistically significant differences across multiple evaluation metrics - including report quality scores, pairwise preference testing, consistency assessments, and reading time - the study plans to enroll 300 patients.

**Supplementary Figure 1** Receiver operating characteristic (ROC) curves (AUC<0.8) of 14 predictive features in the CXR-27 Test Set.

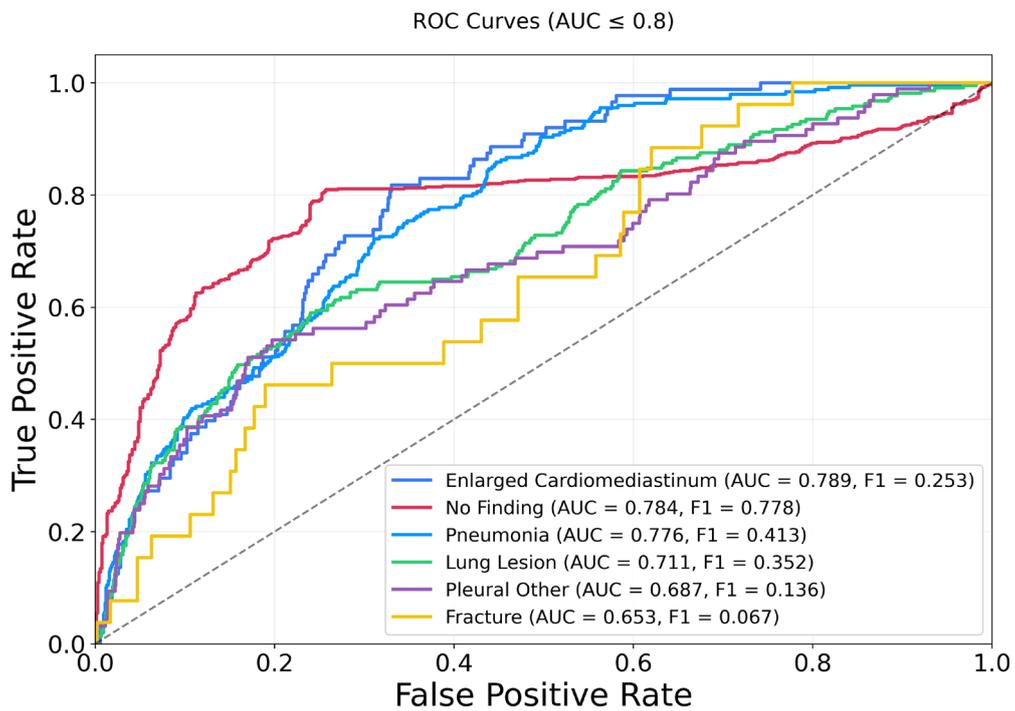

The ROC curves were constructed using probability thresholds ranging continuously from 0 to 1. Notably, the edema classification was excluded from analysis due to insufficient positive cases (n=1), as such limited sample size would yield non-representative ROC curve characteristics.

**Supplementary Figure 2** Confusion matrix evaluation of large model-generated radiology reports

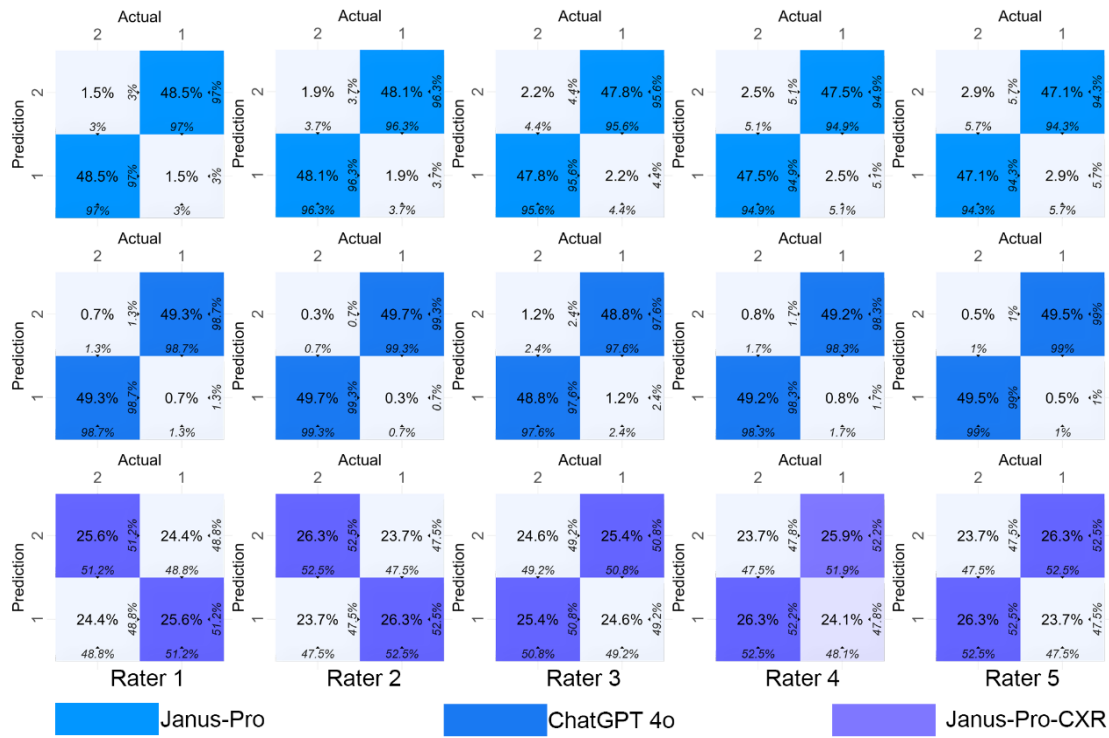

Five experts independently evaluated the style and formatting of AI-generated reports by comparing them against published reference reports, while deliberately excluding consideration of content accuracy.

**Supplementary Figure 3** Interactive Interface for Professional Medical Imaging Analysis

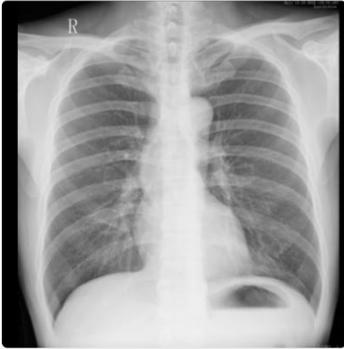

Users can switch between different models and request a detailed report on a chest X - ray image from the perspective of a radiology expert in the "Question" box, including FINDINGS and IMPRESSION. The "Answer" box provides the analysis results in English, and the "Translation" box offers the corresponding Chinese translation. The chest X - ray image is displayed on the right side of the interface.

**Supplementary Table 1** Baseline characteristics of retrospective study patients.

| Characteristic | Fine-tuning | | Testing | |
| --- | --- | --- | --- | --- |
| | MIMIC-CXR (n=220000) | CXR-27 (n=5267) | MIMIC-CXR (n=1317) | CXR-27 (n=1471) |
| Age, years (mean ± s.d.) | - | 41.7±22.4 | 39.2±28.1 | 37.9±28.0 |
| Age class, n (%) | | | | |
| <18 | - | 890 (16.9%) | 192 (14.6%) | 259 (17.6%) |
| 18-60 | - | 3123 (59.3%) | 703 (53.4%) | 780 (53.0%) |
| >60 | - | 1254 (23.8%) | 422 (32.0%) | 432 (29.4%) |
| Sex, n (%) | | | | |
| male | - | 2944 (55.9%) | 721 (54.7%) | 723 (49.2%) |
| female | - | 2322 (44.1%) | 595 (45.2%) | 747 (50.8%) |
| Disease, n (%) | | | | |
| No Finding | 75163 (33.0%) | 2393 (45.4%) | 641 (48.7%) | 101 (6.9%) |
| Support Devices | 65637 (28.8%) | 547 (10.4%) | 117 (8.9%) | 375 (25.5%) |
| Pleural Effusion | 1961 (0.9%) | 538 (10.2%) | 127 (9.6%) | 371 (25.2%) |
| Lung Opacity | 50916 (22.3%) | 2018 (38.3%) | 486 (36.9%) | 508 (34.5%) |
| Pneumonia | 15769 (6.9%) | 1100 (20.9%) | 248 (18.8%) | 99 (6.7%) |
| Cardiomegaly | 39094 (17.2%) | 494 (9.4%) | 109 (8.3%) | 574 (39.0%) |
| Lung Lesion | 6129 (2.7%) | 862 (16.4%) | 217 (16.5%) | 93 (6.3%) |
| Enlarged Cardiomediastinum | 7004 (3.1%) | 369 (7.0%) | 88 (6.7%) | 123 (8.4%) |
| Pneumothorax | 9317 (4.1%) | 254 (4.8%) | 58 (4.4%) | 14 (1.0%) |
| Atelectasis | 45088 (19.8%) | 355 (6.7%) | 75 (5.7%) | 356 (24.2%) |
| Pleural Other | 1961 (0.9%) | 460 (8.7%) | 96 (7.3%) | 51 (3.5%) |
| Consolidation | 10487 (4.6%) | 83 (1.6%) | 16 (1.2%) | 78 (5.3%) |
| Fracture | 3768 (1.7%) | 96 (1.8%) | 26 (2.0%) | 88 (6.0%) |
| Edema | 26455 (11.6%) | 3 (0.1%) | 1 (0.1%) | 266 (18.1%) |

The retrospective data were from 27 medical centers in China and were named CXR-27. These data were divided into a fine-tuning set and a test set at a ratio of 8:2.

**Supplementary Table 2** Baseline characteristics of prospective study participants.

| Characteristic | Value (n = 297) |
| --- | --- |
| Age, years (mean ± s.d.) | 39.2±26.0 |
| Age class, n (%) | |
|   <18 | 90 (30.3%) |
|   18-60 | 123 (41.4%) |
|   >60 | 84 (28.3%) |
| Sex, n (%) | |
|   male | 160 (53.9%) |
|   female | 137 (46.1%) |
| Disease, n (%) | |
|   No Finding | 91 (30.6%) |
|   Support Devices | 73 (24.6%) |
|   Pleural Effusion | 56 (18.9%) |
|   Lung Opacity | 169 (56.9%) |
|   Pneumonia | 129 (43.4%) |
|   Cardiomegaly | 60 (20.2%) |
|   Lung Lesion | 73 (24.6%) |
|   Enlarged Cardiomediastinum | 58 (19.5%) |
|   Pneumothorax | 14 (4.7%) |
|   Atelectasis | 11 (3.7%) |
|   Pleural Other | 38 (12.8%) |
|   Consolidation | 50 (16.8%) |
|   Fracture | 10 (3.4%) |
|   Edema | 0 (0%) |

While the CXR-27 dataset naturally contains approximately 50% 'no finding' cases, we prospectively reduced this prevalence to 30% to construct a more diagnostically challenging evaluation dataset. This deliberate prevalence adjustment was designed to rigorously test the AI system's capability in identifying true abnormalities under conditions of reduced negative cases.

**Supplementary Table 3** Automated report generation metrics.

(A) Automated report generation metrics on MIMIC-CXR Test Set.

| Model | Year | Size | CheXbert label | | | | Rad-Graph | BLEU | | ROUGE |
|---|---|---|---|---|---|---|---|---|---|---|
| | | | Micro-avg | | Macro-avg | | | | | |
| | | | F1-14 | F1-5 | F1-14 | F1-5 | F1 | -1 | -4 | -L |
| Janus-CXR(OURS)* | 2025 | 1B | <u>47.3</u> | <u>57.2</u> | **34.7** | **50.1** | **26.4** | 33 | 11 | 28.6 |
| GPT-4V | \ | \ | 35.6 | 33.3 | 25.3 | 29.6 | 13.2 | 16 | 1.9 | 13.2 |
| HealthGPT*[3] | 2025 | 3.8B | 22.6 | 24.6 | 16.5 | 20.1 | 14.1 | 18 | 1.2 | 14.1 |
| FlamingoCXR[4] | 2024 | 3B | **51.9** | **58** | \ | \ | 20.5 | \ | 10 | **29.7** |
| PromptMRG*[5] | 2024 | <1B | 15 | 6.9 | 8.44 | 4.09 | 20 | <u>40</u> | 11 | 26.8 |
| Med-PaLM M[6] | 2023 | 12B | \ | \ | \ | \ | <u>25.2</u> | 31 | 10 | 26.2 |
| LLaVA-Med[7] | 2023 | 7B | 27.3 | 24.4 | 18.7 | 20.5 | 6.5 | 21 | 1.3 | 13.8 |
| CvT-21DistillGPT2*[8] | 2023 | <1B | 41 | 46.3 | 28.2 | 41 | 23.8 | 39 | <u>13</u> | 28.6 |
| RGRG*[9] | 2023 | <1B | 37.4 | 49 | 24.4 | 42.7 | 23.9 | 37 | 13 | 26.4 |
| R2GenGPT[10] | 2023 | 7B | 38.9 | \ | \ | \ | \ | **41** | 13 | **29.7** |
| BioVil-T[11] | 2023 | <1B | 31.7 | \ | \ | \ | \ | \ | 9.1 | 29.6 |

(B) Automated report generation metrics on CXR-27 Test Set.

| Model | Year | Size | DeepSeek label | | | | Rad-Graph | BLEU | | ROUGE |
|---|---|---|---|---|---|---|---|---|---|---|
| | | | Micro-avg | | Macro-avg | | | | | |
| | | | F1-14 | F1-5* | F1-14 | F1-5* | F1 | -1 | -4 | -L |
| Janus-Pro-CXR-400 (Ours) | 2025 | 1B | **56.2** | **54.5** | **35.1** | **55.1** | **61.1** | **62** | **44** | **60.5** |
| Janus-Pro-CXR-zero (Ours) | 2025 | 1B | <u>46</u> | <u>27.9</u> | <u>25.2</u> | <u>29.2</u> | 16.1 | 18 | 3.8 | <u>22</u> |
| HealthGPT[3] | 2025 | 3.8B | 37.9 | <u>27.9</u> | 13.6 | 18.5 | <u>21.1</u> | <u>30</u> | <u>5.1</u> | 19.2 |
| PromptMRG[5] | 2024 | <1B | 36.2 | 3 | 8.5 | 4.3 | 10.8 | 21 | 1.8 | 16.7 |
| CvT-21DistillGPT2[8] | 2023 | <1B | 42.4 | 17.7 | 19.7 | 15.6 | 8.9 | 10 | 1.2 | 17.2 |
| RGRG[9] | 2023 | <1B | 37.6 | 20 | 18.6 | 16.6 | 13.2 | 28 | 2.9 | 20.2 |

(A) Automated report generation metrics on the MIMIC test set, with data sourced from the MIMIC-CXR dataset. CheXbert labeling tool was used for annotation (uncertain labels were treated as positive). * indicates open-source models - these large models were tested (with parameter configurations strictly following the original papers), while data for non-open-source models were obtained from their respective publications. Top 5 conditions in MIMIC test set: Atelectasis, Cardiomegaly, Edema, Consolidation, Pleural Effusion. The bold represents the highest value, and the underline represents the second highest value.

(B) Automated report generation metrics on the CXR-27 test set, with data sourced from retrospective studies. DeepSeek labeling tool was employed, testing the same open-source large models (parameter configurations strictly followed original papers). Top 5 conditions in CXR-27 test set: Support Devices, Pleural Effusion, Lung Opacity, Pneumonia, Lung Lesion. The top performer is shown in bold, while the runner-up is underlined. F1-14 represents F1 scores across 14 diseases; F1-5 covers the five most common conditions. B denotes model parameters in billions (1B = 1 billion parameters), where typically fewer parameters indicate lower model development and operational costs. The bold represents the highest value, and the underline represents the second highest value.

**Supplementary Table 4** Model performance evaluation using F1 scores on the CXR-27 test set

| Model | F1 | | | | | | | | | | | | | |
|---|---|---|---|---|---|---|---|---|---|---|---|---|---|---|
| | ECm. | Cmgl. | Opac. | Les. | Edema | Cnsl. | Pna. | Atel. | Pmtx. | Eff. | P.O. | Frac. | Dev. | NoF. |
| Janus-Pro-CXR-400 (Ours) | **25.3** | 31.0 | **60.2** | **35.2** | 0.0 | 10.5 | **41.3** | 22.3 | 26.8 | **65.2** | **13.6** | **6.7** | **73.6** | **77.8** |
| Janus-Pro-CXR-zero (Ours) | 11.8 | 25.2 | 30.0 | 8.6 | 0.0 | **25.0** | 15.4 | **35.9** | **36.0** | 62.7 | 2.0 | 3.6 | 23.7 | 72.9 |
| HealthGPT[3] | 8.5 | 0.0 | 44.9 | 22.1 | 0.0 | 5.7 | 3.2 | 16.5 | 0.0 | 20.5 | 1.9 | 0.0 | 1.6 | 65.0 |
| PromptMRG[5] | 12.1 | 3.6 | 0.0 | 0.0 | 0.0 | 0.0 | 0.0 | 15.5 | 0.0 | 0.0 | 0.0 | 0.0 | 21.2 | 66.9 |
| CvT-21DistillGPT2[8] | 25.0 | 30.6 | 26.5 | 0.9 | 0.0 | 19.5 | 6.8 | 29.6 | 17.9 | 28.8 | 5.9 | 0.0 | 14.8 | 69.8 |
| RGRG[9] | 0.0 | **41.2** | 28.4 | 0.0 | 0.0 | 19.2 | 19.5 | 21.8 | 25.2 | 25.2 | 0.0 | 0.0 | 10.1 | 69.9 |

The open-source large models were tested with parameter configurations strictly adhering to the original papers, where the top-performing model is presented in **bold** and the second-best model is indicated with *underscores*. ECm., Enlarged Cardiomediastinum; Cmgl., Cardiomegaly; Opac., Lung Opacity; Les., Lung Lesion; Cnsl., Consolidation; Pna., Pneumonia; Atel., Atelectasis; Pmtx., Pneumothorax; Eff., Pleural Effusion; P.O., Pleural Other; Frac., Fracture; Dev., Support Devices; NoF., No Finding.

**Supplementary Table 5** A Example of AI improving junior radiologists' reporting accuracy.

| Image | Report by senior radiologists reviewed and released | Report by a junior radiologist (SCP group) | AI-generated report | AI-assisted report |
|---|---|---|---|---|
| 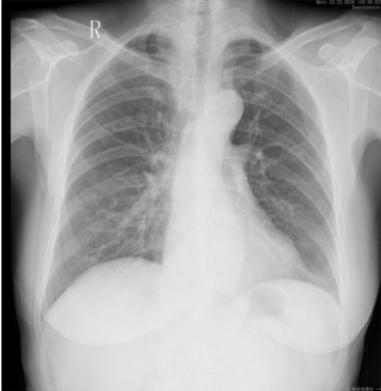 | **Findings:** The thorax is symmetrical, and the trachea is centrally located. The lung markings are increased bilaterally, with no enlargement of the hilar shadows. The aorta is tortuous, while the size and shape of the heart appear normal. Both diaphragmatic surfaces are smooth, and the costophrenic angles are sharp. **Impression:** 1. Tortuous aorta. 2. Bronchitis. Further CT examination may be considered if necessary. | **Findings:** The lung fields are clear bilaterally with no significant parenchymal lesions noted. The trachea and mediastinum are midline. The cardiac silhouette is not enlarged. The aorta is tortuous. The diaphragmatic contours are smooth bilaterally, and the costophrenic angles are sharp. **Impression:** Tortuosity of the aorta. | **Findings:** Increased lung markings in both lungs with partial blurring of margins. No significant cardiac enlargement. Tortuous aorta with calcification. Smooth diaphragmatic surfaces and sharp costophrenic angles bilaterally. **Impression:** 1. ==Bronchitis with minor inflammatory changes in both lungs.== 2. Tortuous aorta with calcification. Clinical correlation is recommended, and further CT examination may be considered if necessary. | **Findings:** Increased lung markings with partially blurred margins in both lungs. No significant enlargement of the cardiac silhouette. Tortuous aorta with calcification. Smooth diaphragmatic surfaces and sharp costophrenic angles bilaterally. **Impression:** 1. ==Bronchitis with minor inflammatory changes in both lungs.== 2. Tortuous aorta with calcification. Clinical correlation is recommended, and further CT examination may be considered if necessary. |